\documentclass[a4paper,
              ]{jacow}

\makeatletter
\ifboolexpr{bool{xetex}}                 
 {\renewcommand{\Gin@extensions}{.pdf,%
                    .png,.jpg,.bmp,.pict,.tif,.psd,.mac,.sga,.tga,.gif,%
                    .eps,.ps,%
                    }}{}
\makeatother

\ifboolexpr{bool{xetex} or bool{luatex}} 
 {}                                      
 {\usepackage[utf8]{inputenc}}           

\usepackage[english]{babel}

\ifboolexpr{bool{jacowbiblatex}}
 {%
  \addbibresource{jacow-test.bib}
  \addbibresource{biblatex-examples.bib}
 }{}

\begin{document}

\title{STUDY OF ULTRA-HIGH GRADIENT ACCELERATION IN CARBON NANOTUBE ARRAYS\thanks{Work supported by Cockcroft Institute core Grant No. ST/G008248/1.}}

\author{J. Resta-L\'opez\thanks{jrestalo@liverpool.ac.uk}, A. Alexandrova, V. Rodin, Y. Wei, C. P. Welsch, \\
Cockcroft Institute and The University of Liverpool, UK\\
Y. Li, G. Xia, Y. Zhao, \\
Cockcroft Institute and The University of Manchester, UK}

\maketitle

\begin{abstract}
Solid-state based wakefield acceleration of charged particles was previously proposed to obtain extremely high gradients on the order of 1 -- 10 TeV$/$m. In recent years the possibility of using either metallic or carbon nanotube structures is attracting new attention. The use of carbon nanotubes would allow us to accelerate and channel particles overcoming many of the limitations of using natural crystals, e.g. channeling aperture restrictions and thermal-mechanical robustness issues. In this paper, we propose a potential proof of concept experiment using carbon nanotube arrays, assuming the beam parameters and conditions of accelerator facilities already available, such as CLEAR at CERN and CLARA at Daresbury. The acceleration performance of carbon nanotube arrays is investigated by using a 2D Particle-In-Cell (PIC) model based on a multi-hollow plasma. Optimum experimental beam parameters and system layout are discussed. 
\end{abstract}

\section{introduction}
Solid-state plasma wakefield acceleration using crystals was proposed in the 1980’s and 1990’s \cite{chen1, tajima1, carrigan1, chen2} as an alternative particle acceleration technique to obtain $\sim$TV/m acceleration gradients. However, it has not been experimentally demonstrated yet. In recent years, new efforts have been focused on the feasibility study of channeling acceleration of particle beams using carbon-based nano-crystals such as carbon-nanotubes (CNT) \cite{shin1,shin2} or metallic nanotube structures, e.g. porous alumina \cite{Zhang}. CNT configurations may be advantageous over typical crystal media like silicon because of their large degree of dimensional flexibility and thermo-mechanical strength, which could be suitable for channeling acceleration of MW beams. For example, CNTs allow transverse acceptances of the order of up to 100~nm, i.e. three orders of magnitude higher than a typical silicon channel. Therefore, CNTs might be used for wakefield acceleration using either a beam or a laser as driving source. 

The commissioning of new facilities, such as CLARA/PARS at Daresbury \cite{xia1} and CLEAR at CERN \cite{navarro}, could be a great opportunity to design research stations to explore plasma wakefield concepts and also novel solid state based channeling acceleration concepts. Concretely, here the feasibility of using CNTs for channeling acceleration is investigated by means of simulations using Particle-In-Cell (PIC) software tools. In addition, a potential proof-of-principle experiment is drafted, using the electron beam of CLARA or CLEAR. 


\section{Theoretical background}
The electromagnetic interaction of a charged particle travelling through a CNT can be described by means of a linearized hydrodynamic model for a plasma perturbation \cite{Arista, Stockli, Wang1, Wang2}, assuming a 2D electron gas confined in the cylindrical shell of a CNT. 

Let us consider a point-like charge $q$ moving with ultra-relativistic velocity $v\approx c$ inside a single-walled carbon nanotube (SWNT) on the trajectory parallel to the nanotube axis $z$: ${\bf r}_0=(r_0, \varphi_0, z_0)$, assuming cylindrical coordinates and a nanotube radius $r=a$.  The electronic excitations on the nanotube wall can be described by the continuity equation:

\begin{equation}
\partial_t n_1({\bf r}_a, t) + n_0 \nabla_{\parallel} \cdot {\bf u}({\bf r}_a, t)=0,
\label{eq:1}
\end{equation}

\noindent the momentum-balance equation:

\begin{align}
\partial_t {\bf u} ({\bf r}_a, t) &=  \frac{e}{m_e}\nabla_{\parallel} \Phi({\bf r}_a, t) - \frac{\alpha}{n_0} \nabla_{\parallel} n_1({\bf r}_a, t) \nonumber \\
& -  \gamma {\bf u}({\bf r}_a, t) + \frac{\beta}{n_0} \nabla_{\parallel}\Big[\nabla^2_{\parallel} n_1 ({\bf r}_a, t)\Big],
\label{eq:2}
\end{align}

\noindent and Poisson's equation:
\begin{equation}
\nabla^2 \Phi ({\bf r, t})=\frac{1}{\epsilon_0} \big[e n_1({\bf r}_a, t) - q\delta({\bf r} -{\bf r}_0)\big].
\label{eq:3}
\end{equation}

In the previous equations the perturbed state of the electron fluid in the surface is described by the electron number density per unit area: $n_0 + n_1({\bf r}_a, t)$, where $n_0$ is the unperturbed plasma density and $n_1$ the perturbed plasma density. The vector ${\bf u}$ represents the velocity field of the electron fluid and ${\bf r}_a$ is the vector position at the nanotube surface. 

In Eq.~(\ref{eq:2}), $e$ is the unit electric charge and $m_e$ the electron mass. $\Phi$ is the electric potential which results from the external charge $q$ and the charge-density polarization of the electron gas $n_1$. The second term on the right-hand side of Eq.~(\ref{eq:2}) is the force due to internal pressure in a 2D homogeneous Thomas-Fermi electron fluid with $\alpha=v^2_F/2$ and Fermi velocity $v_F=\hbar/m_e \sqrt{2 \pi n_0}$; the third term is related to the frictional force on electrons due to scattering on the positive charge background with a damping coefficient $\gamma$; and the fourth term with $\beta=1/4$ is a quantum correction. 

Following  \cite{Arista, Stockli, Wang1, Wang2} and taking into account the natural boundary conditions at $r=0$ and $r\rightarrow \infty$, the potential components can be expanded in terms of cylindrical Bessel functions $I_m(x)$ and $K_m(x)$ of integer order $m$. The total potential inside the nanotube can be calculated as 
$\Phi = \Phi_0 + \Phi_{ind}$, where $\Phi_0$ is the Coulomb potential contribution from the driving charge:

\begin{align}
\Phi_{0}(r,\varphi, z, t)  & =  \frac{1}{4\pi \epsilon_0}  \frac{q}{\lVert {\bf r} - {\bf r_0}\rVert} \nonumber \\
{} & =  \frac{q}{4 \pi^2 \epsilon_0} \sum_{m=-\infty}^{+\infty} \int_{-\infty}^{+\infty} \mathrm{d}k e^{ik\zeta+im(\varphi - \varphi_0)} \nonumber \\
&   \times I_m(|k|r_{<}) K_m(|k|r_{>}),
\label{eq:4}
\end{align}

\noindent where $r_{<}$ ($r_{>}$) is the smaller (larger) of $r$ and $r_0$, and $k$ the plasma longitudinal wave number. $\zeta=z-ct$ is the co-moving coordinate.

$\Phi_{ind}$ is the induced potential due to the perturbation of the electron fluid on the carbon nanotube surface:

\begin{align}
\Phi_{ind}(r,\varphi, z, t)  & =   \frac{q}{4 \pi^2 \epsilon_0} \sum_{m=-\infty}^{+\infty} \int_{-\infty}^{+\infty} \mathrm{d}k e^{ik\zeta+im(\varphi - \varphi_0)} \nonumber \\
&  \times I_m(|k|r_0) I_m(|k|r) A_m(k),
\label{eq:5} 
\end{align}

\noindent where $A_m(k)$ is a non-dimensional function given by:

\begin{equation}
A_m(k)  = \frac{\Omega^2_p a^2 (k^2 + m^2/a^2) K^2_m(|k| a)}{kc(kc + i\gamma) - \omega^2_m(k)}, 
\label{eq:6}
\end{equation}

\noindent with the resonant frequency:

\begin{align}
\omega^2_m(k) & =  \alpha (k^2+m^2/a^2) + \beta(k^2 + m^2/a^2)^2 \nonumber \\
& + \Omega^2_p a^2 (k^2 + m^2/a^2) K_m(|k|a) I_m(|k|a),
\label{eq:7}
\end{align}

\noindent with $\Omega_p= \sqrt{e^2 n_0/(\epsilon_0 m_e a)}$.

Then the corresponding longitudinal wakefield can be calculated as $W_{z}=  -\partial_{\zeta} \Phi$,

\begin{align}
W_{z}(r,\varphi,z,t) & =  \frac{q}{4\pi^2 \epsilon_0}  \sum_{m=-\infty}^{+\infty} \int_{-\infty}^{+\infty} \mathrm{d}k k e^{im(\varphi - \varphi_0)} \nonumber \\
{} &  \times \Big[ I_m(|k|r_{<}) K_m(|k|r_{>}) \sin (k\zeta) \nonumber \\
{} &  +  I_m(|k| r_0) I_m(|k| r) \Big( \Re [A_m(k)] \sin(k\zeta) \nonumber \\
{} &  +  \Im [A_m(k)]\cos(k\zeta) \Big) \Big].
\end{align}

The longitudinal wake of a point-like charge $W_{z}$ can be used as a Green's function to compute the longitudinal wakefield generated by a driving bunch with arbitrary charge distribution $\rho (z)$:

\begin{equation}
E_{z}(\zeta)=\int_{\zeta}^{\infty} \mathrm{d}\zeta'  \rho(\zeta') W_{z}(\zeta-\zeta').
\label{eq:9}
\end{equation}

\section{Simulations}

\begin{figure}[!tbh]
    \centering
    \includegraphics*[width=5cm]{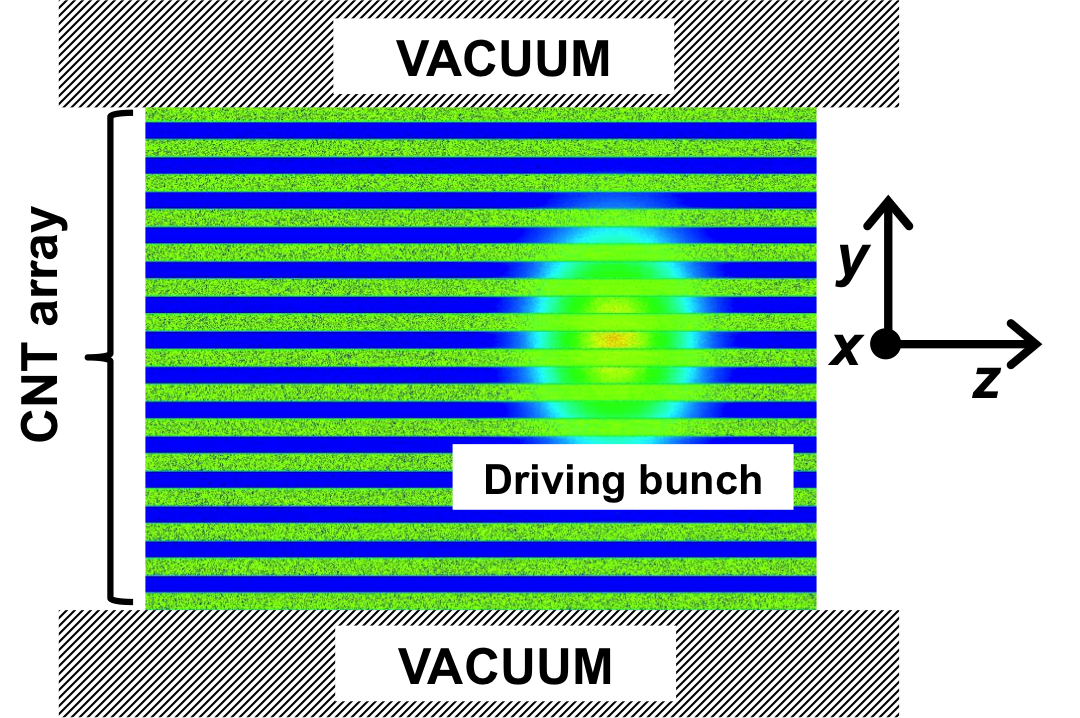}
    \caption{Schematic of the simulation model: alternating hollow channels and plasma walls inside a vacuum chamber.}
    \label{fig:schematicmodel}
\end{figure}

With the open source PIC code EPOCH \cite{epoch}, we set up the 2D plane geometry of a CNT array to explore its interaction with a bi-gaussian driving bunch (Fig.~\ref{fig:schematicmodel}). The array is modelled by a multi-hollow plasma where high density plasma layers act as the nanotube walls and provide electron fluids. In reality, the electron fluids could come from the plasmon excitation by external electromagnetic fields from driving beams or lasers. The electron density in solid-state plasmas is normally in the range of $10^{25}$~m$^{-3}< n_e < 10^{30}~$m$^{-3}$ \cite{chen1, Ostling}. To better accommodate the dimensions of realistic bunches, we adopt a relatively low wall plasma density of $10^{25}~$m$^{-3}$ in simulations. 


\begin{figure}[!b]
    \centering
    \includegraphics*[width=8cm]{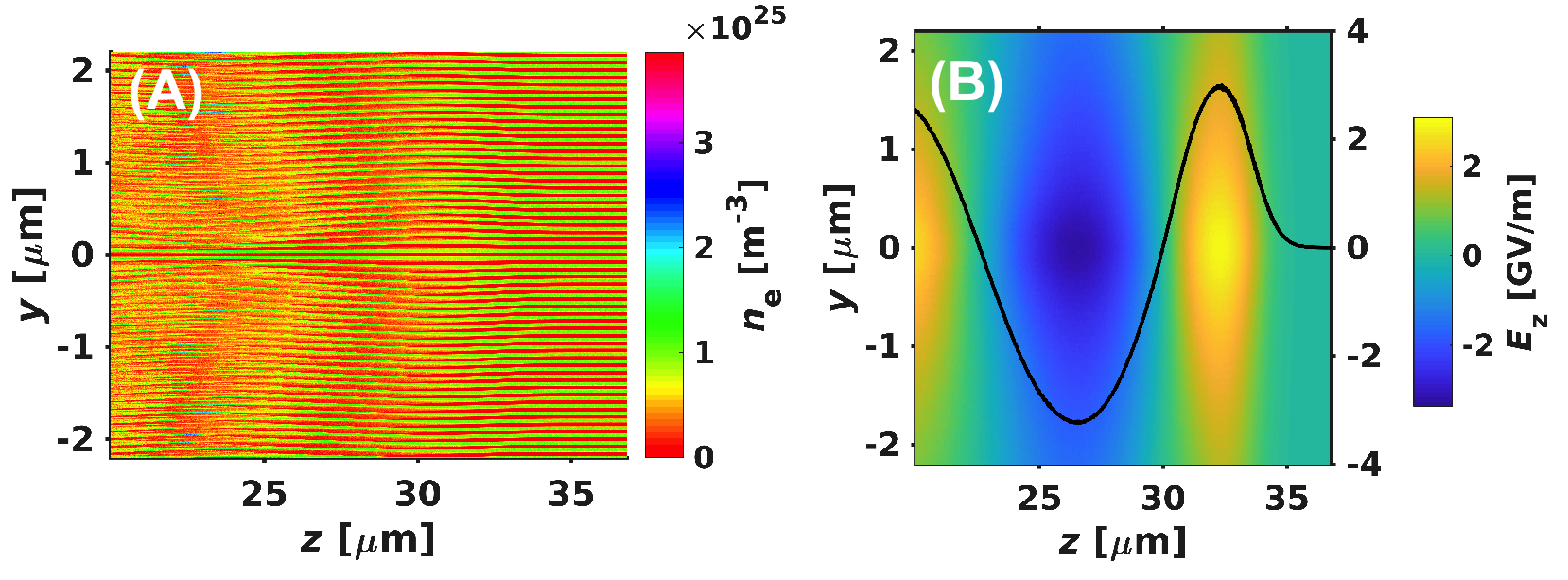} 
    \caption{(A) Plasma electron density perturbation and (B) longitudinal electric field at the propagation distance $z=20~\mu$m for the case $n_b/n_e=0.1$.}
    \label{fig:gradient2}
\end{figure}

\begin{figure}[!t]
    \centering
    \includegraphics*[width=8cm]{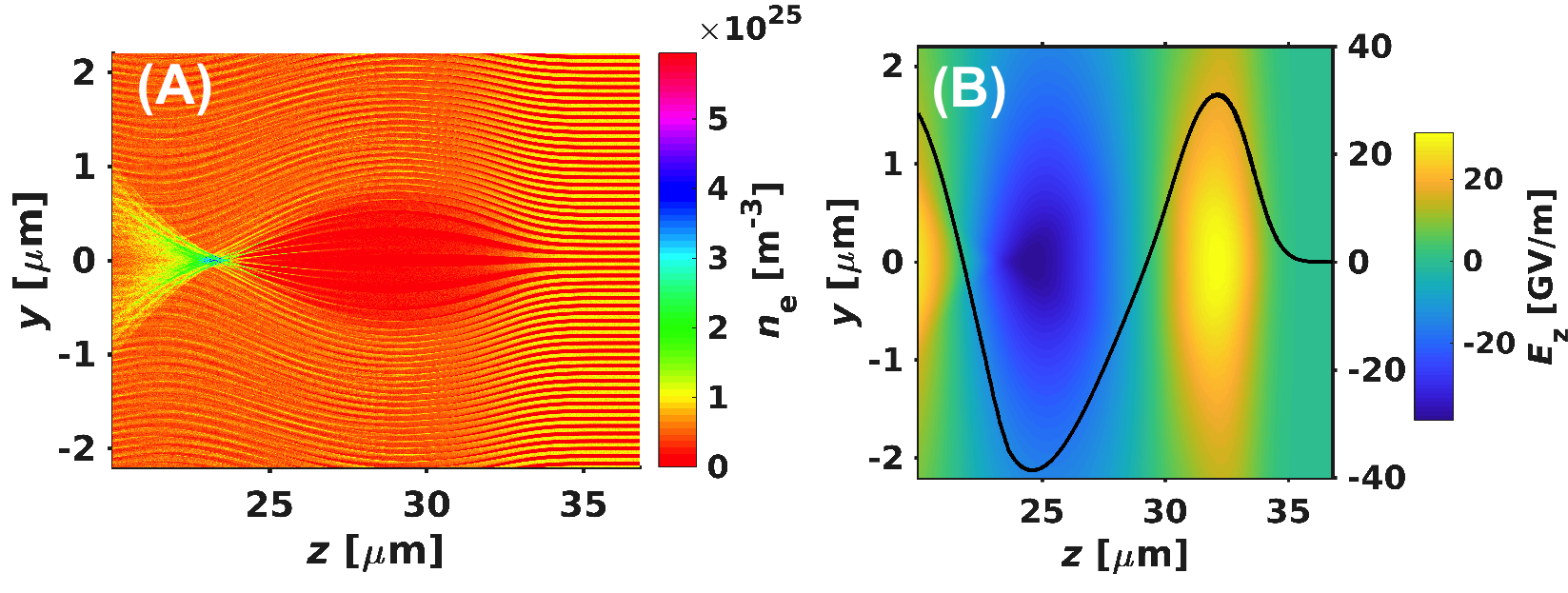}
    \caption{(A) Plasma electron density perturbation and (B) longitudinal electric field at the propagation distance $z=20~\mu$m for the case $n_b/n_e=1$.}
    \label{fig:gradient1}
\end{figure}

The CNT channel radius is 20 nm and the plasma wall thickness is 40 nm. The CNT array is wide enough so that there are unperturbed tubes at the boundaries. The electron bunch propagating through the CNT array is initialized as follows: rms bunch length $\sigma_z=0.5 c/\omega_p$, rms bunch radius $\sigma_r=0.1 c/\omega_p$, bunch energy $W=200$ MeV and energy spread $\delta W/W=1\%$, where $\omega_p=\sqrt{n_e e^2/(m_e \epsilon_0)}$ is the angular plasma frequency of the wall and $c$ is the speed of light. The simulation variant is the bunch population (i.e., bunch density $n_b$). When $n_b=0.1 n_e$, perturbation of the plasma electrons is moderate and the tube structure is still observable (Fig.~\ref{fig:gradient2}). Therefore, relatively low electric fields are excited under weak beam-plasma coupling. The accelerating gradient is around $3$~GV$/$m. With higher beam density ($n_b=n_e$), we see a clear plasma density perturbation (Fig.~\ref{fig:gradient1}). Also the plasma electrons follow similar trajectories as in a uniform plasma. The accelerating gradient reaches up to $40$ GV$/$m. 

\begin{figure}[!tbh]
    \centering
    \includegraphics*[width=5.5cm]{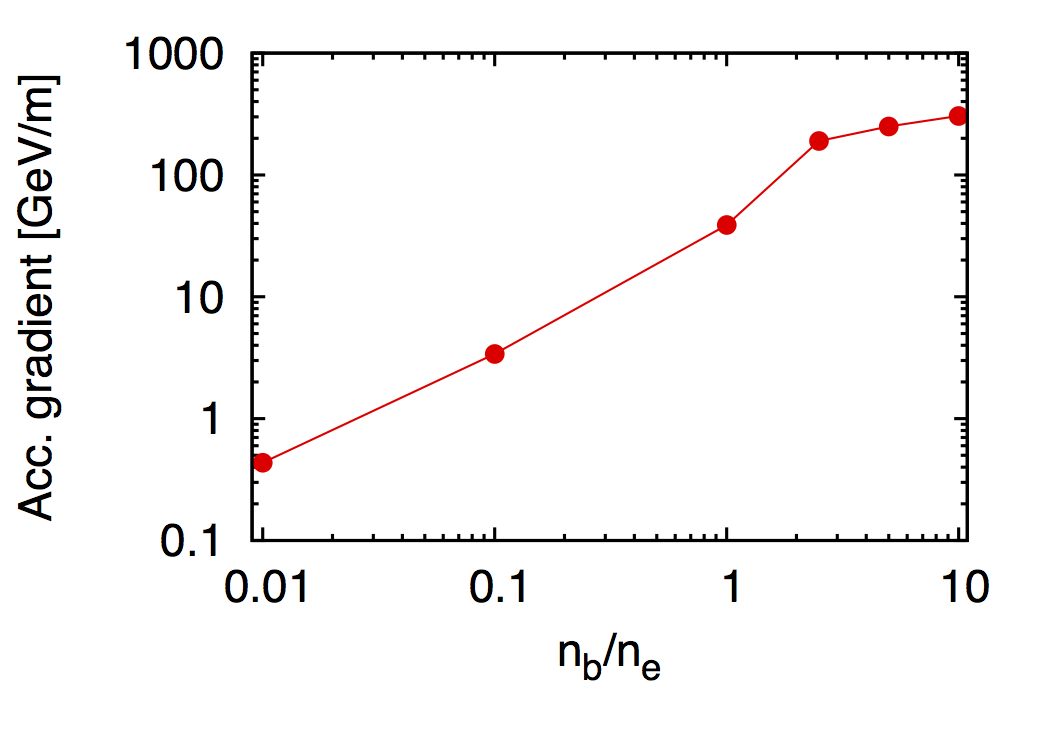}
    \caption{Acceleration gradient as a function of driving bunch density.}
    \label{fig:densitychange}
\end{figure}

Figure~\ref{fig:densitychange} depicts the acceleration gradient as a function of the driving bunch density normalized by the wall plasma density of $n_e=10^{25}~$m$^{-3}$. Similar to the uniform plasma case, a larger beam density can drive stronger plasma wakefields. When the beam density reaches almost 2.5 times the plasma density, the beam-plasma interaction transfers from the linear to the nonlinear regime. The wave breaks when the beam density increases further and plasma electrons get trapped into the wakefields. The accelerating gradient reaches the wave-breaking limit at $E_z=m_e c \omega_p/e \simeq 305$~GV/m. Apart from the bunch density, the tube parameters affect the wake excitation significantly as well. For instance, the increase of plasma wall thickness (e.g., using multi-walled nanotubes) and decrease of the tube radius can enhance the wakefields. This suggests that further optimization of the tube arrays is necessary.

With the same beam and plasma density, we further examine the beam dynamics of the driver in the uniform plasma. Figure 5 shows that more electrons are confined in a smaller core and obtain smaller transverse momenta after propagating in the multi-hollow plasma structure, while the uniform plasma scatters the electrons widely. This suggests that the CNT array helps to efficiently cool the transverse phase space of channelled beams.


\begin{figure}[!tbh]
    \centering
    \includegraphics*[width=8cm]{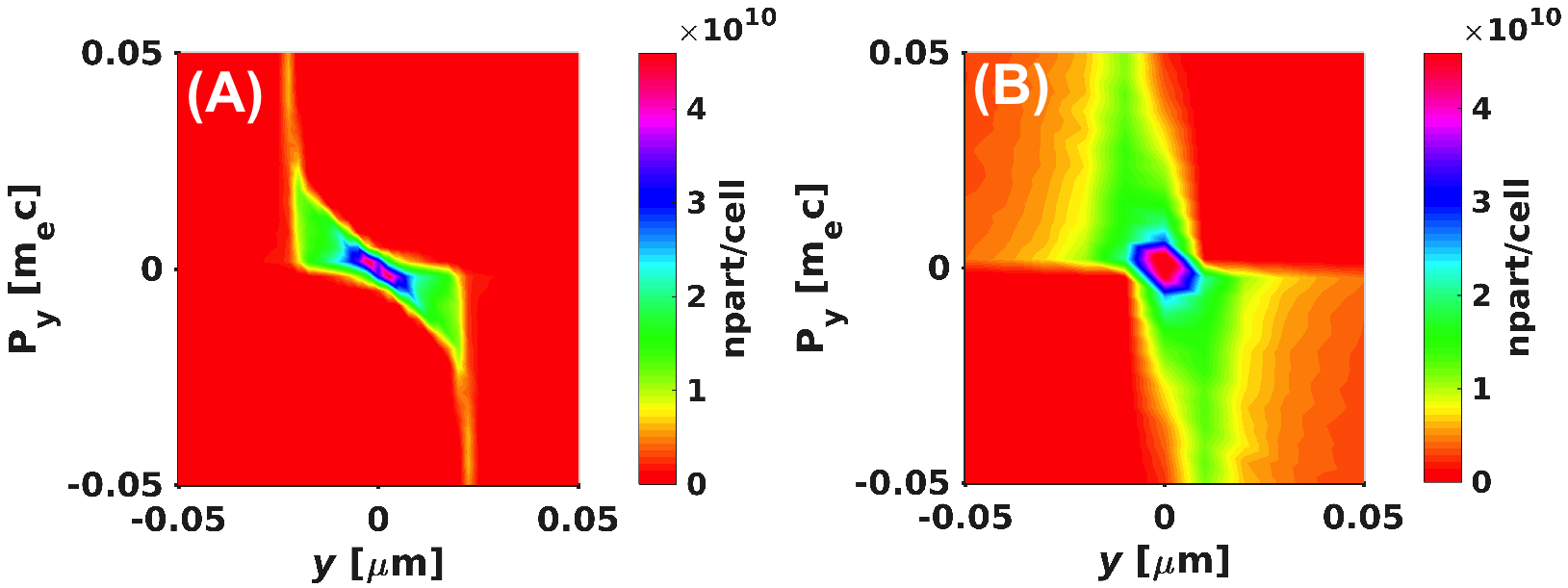}
    \caption{Transverse phase space of the driver after propagating for 50~$\mu$m in the CNT array (A) and uniform plasma with the same plasma density (B) for $n_b=n_e$.}
    \label{fig:transv}
\end{figure}

\section{Experimental layout}
\begin{figure}
\centering
\includegraphics*[width=8cm]{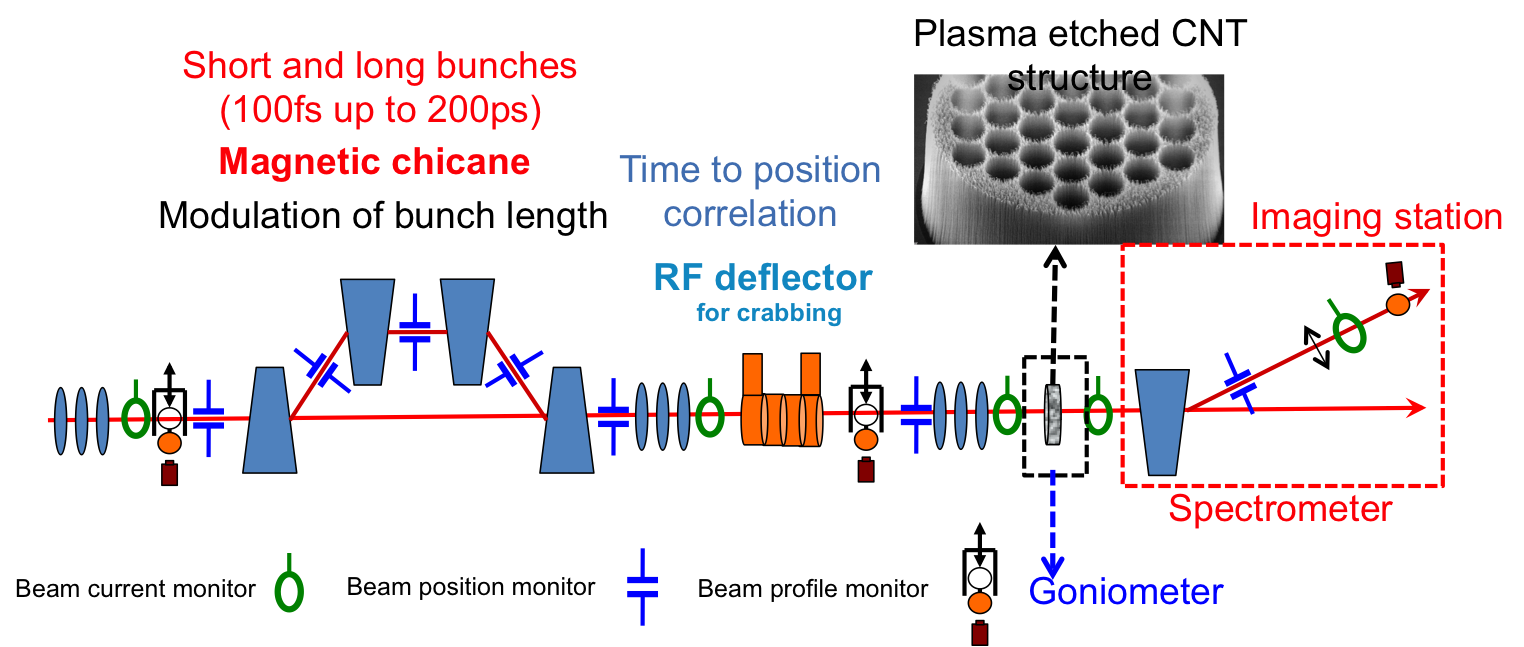}
\caption{Layout of the proposed experimental beamline for the CNT based wakefield acceleration test.}
\label{fig:explayout}
\end{figure}

The experimental setup is shown in Fig.~\ref{fig:explayout}. Such an experimental beamline might be feasible in accelerator facilities such as CLEAR at CERN or CLARA at Daresbury. They will operate in a similar range of energies, $\sim 200$~MeV. In both cases we expect to operate with short bunches on the order of 0.1 ps, and the beam can be modulated by a bunch compresor chicane. If necessary, even shorter bunches could be obtained at the sub-fs level via bunch slicing in the magnetic chicane, using a collimator \cite{Assmann}. 





Taking into account the beam dimensions and charge range in CLEAR and CLARA, the experiment will likely operate in the linear regime $n_b \ll n_e$, and $E_z \propto n_b$. For instance, if the driving beam parameters are matched to obtain a challenging value of $n_b/n_e \sim 0.001$, then $E_z \approx 40$~MV/m from extrapolation in Fig.~\ref{fig:densitychange}. Assuming a CNT array length of 1~mm, then the energy gain is $\Delta W \approx 40$~keV. Therefore, a spectrometer resolution of $\lesssim 10$~keV will be required. 

\section{outlook}
The use of solid nano-structures may open new possibilities to obtain high particle acceleration gradients beyond those provided by standard RF technologies. Amongst them, a CNT array could offer the following advantages with respect to conventional gaseous plasmas: emittance damping through channeling and higher plasmon density. CNT structures also have a higher thermal and mechanical robustness with respect to other solid structures, such as metallic crystals. 

To explore the capabilities of wakefield channeling acceleration using CNT arrays, we have used a 2D PIC multi-hollow plasma model. Here we have considered plasmon excitations by a driving bunch. Preliminary results show the possibility of obtaining longitudinal electric acceleration gradients $> 10$~GV/m. 

New test beam facilities, such as CLEAR and CLARA might offer the opportunity to carry out a proof-of-concept test of CNT based wakefield acceleration. Studies are in progress to address several challenges.





\end{document}